\documentclass{ws-mpla}
\usepackage[super,compress]{cite}
\usepackage{graphicx}

\usepackage{amssymb}
\usepackage{lipsum}
\usepackage{epsfig,graphics}
\usepackage{graphicx}
\usepackage{amsmath}
\usepackage{xcolor}

\begin{document}
\markboth{Sahoo et. al.}{Examining the influence of hadronic interactions on the directed flow ...}
%%%%%%%%%%%%%%%%%%%%% Publisher's Area please ignore %%%%%%%%%%%%%%
\catchline{}{}{}{}{}
%%%%%%%%%%%%%%%%%%%%%%%%%%%%%%%%%%%%%%%%%%%%%%%%%%%%%%%%%%%%%%%%%%%

\title{Examining the influence of hadronic interactions on the directed flow of identified particles in RHIC Beam Energy Scan energies using UrQMD model}

\author{Aswini Kumar Sahoo}\author{Prabhupada Dixit}\author{Md Nasim}
\address{Department of Physical Sciences, Indian Institute of Science Education and Research,\\ Berhampur, Odisha, India
}
\author{Subhash Singha}
\address{Quark Matter Research Center, Institute of Modern Physics, Chinese Academy of Sciences,\\ Lanzhou, Gansu, China}

\maketitle

\pub{Received (Day Month Year)}{Revised (Day Month Year)}

\begin{abstract}
%% Text of abstract
The directed flow of identified particles can serve as a sensitive tool for investigating the interactions during initial and final states in heavy ion collisions. This study examines the rapidity distribution ($dN/dy$), rapidity-odd directed flow ($v_{1}$) and its slope ($dv_{1}/dy$) for $\pi^{\pm}$, $K^{\pm}$, p, and $\bar{\mathrm p}$ in Au+Au collisions at different collision centralities and beam energies ($\sqrt{s_{\mathrm NN}}$ = 7.7, 11.5, 14.5, 19.6, 27, and 39 GeV) using the UrQMD model. We investigate the impact of late-stage hadronic interactions on charge dependent $v_{1}(y)$ and its slope by modifying the duration of the hadronic cascade lifetime ($\tau$). The energy dependence of $dv_{1}/dy$ for p ($\bar{\mathrm p}$) exhibits distinct pattern compared to $\pi^{\pm}$ and $K^{\pm}$. Notably, we observe a change in the sign reversal position of proton $dv_{1}/dy$ at different beam energies with varying $\tau$ in central and mid-central collisions. Moreover, the difference in $dv_{1}/dy$ between positively and negatively charged hadrons ($\Delta dv_{1}/dy$) demonstrates a stark centrality dependence for different particle species. The deuteron displays a significant increase in $dv_{1}/dy$ with increasing $\tau$ compared to p and n. This investigation underscores the importance of considering the temporal evolution and duration of the hadronic phase when interpreting the sign reversal, charge splitting of $v_{1}$ and light nuclei formation at lower RHIC energies.


\end{abstract}

\ccode{PACS}

\section{Introduction}

Extensive research conducted at the Relativistic Heavy Ion Collider (RHIC) and the Large Hadron Collider (LHC) facilities, has exhibited the presence of a fluid state of strongly interacting quark-gluon plasma (QGP)\cite{Adams:2005dq, Adcox:2004mh, Muller:2012zq,BRAHMS:2004adc} in relativistic heavy-ion collisions. An important tool for uncovering the collective behavior in these collisions is azimuthal anisotropic flow coefficients. In such a phenomenon the initial geometric anisotropies are transformed into final momentum space anisotropies of particles due to the expansion of the medium\cite{Ollitrault:1992bk,Poskanzer:1998yz}. This anisotropy can be quantified using a Fourier expansion: 
\begin{equation}
\frac{dN}{d\phi} \propto 1+ 2\sum_{n=1}^{\infty} v_{n} \cos n(\phi - \Psi_{n}), 
\end{equation}

where $v_n$ represents the magnitude of the $n$-th order flow and $\Psi_n$ is the corresponding phase. The first three components of this expansion, namely $v_1$ (directed flow)\cite{Brachmann:1999xt}, $v_2$ (elliptic flow)\cite{Voloshin:1994mz}, and $v_3$ (triangular flow)\cite{Alver:2010gr}, are extensively studied due to their profound physical implications. While $v_2$ and $v_3$ provide insights into the collective expansion within the transverse plane, $v_1$ characterizes the longitudinal evolution by measuring the tilt of the system with respect to the beam direction. However, interpreting $v_1$ is intricate, as its origin involves both global momentum conservation and event-by-event fluctuations~\cite{Luzum:2010fb, Teaney:2010vd, Retinskaya:2012ky}. It is now understood that directed flow comprises two constituents: an odd function in rapidity and an even function in rapidity, with the latter is not related to reaction plane and sensitive to initial fluctuations. 

This paper primarily focuses on the odd-component of directed flow, denoted as $v_{1}$. Over the past decades, extensive research has been conducted on $v_{1}$ across a wide range of beam energies, spanning from the AGS~\cite{E895:2000sor,Chung:2001je}, SPS~\cite{NA49:2003njx}, RHIC~\cite{STAR:2003xyj,PHOBOS:2005ylx}, to LHC~\cite{ALICE:2013xri} energies. Of particular interest is the observation of a change in sign for the directed flow of protons since some model calculations suggested that it could be sensitive to the softening of the Equation of State (EoS) associated with the QCD phase transition~\cite{Stoecker:2004qu,Nara:2016phs,Konchakovski:2014gda}. Indeed, the proton $v_{1}(y)$ measured by the STAR experiment in the RHIC Beam Energy Scan program shows a sign change~\cite{STAR:2014clz,STAR:2017okv} around $\sqrt{s_{\mathrm NN}}$ of 10--20 GeV. It prompted many discussion on the underlying physics of directed flow. However, it has been pointed in ~\cite{Singha:2016mna} that there exists a notable inconsistency among several models that treat the equation of state similarly. Furthermore, even models that do not incorporate the concept of softening are capable of predicting a sign change. Please note that at lower RHIC energies, where one anticipates a baryon-rich environment, the effect of various mean-field potentials on collective flow becomes crucial~\cite{Pal:2000yc,Isse:2005nk,Bravina:2019efl}. However, it is noted in~Ref.~\cite{Nara:2016hbg} that incorporating a hadronic mean field interaction can not account for the negative slope of the proton directed flow. While certain models manage to capture qualitative aspects of the $v_{1}$, they struggle to describe it in a quantitative manner.

An important feature during the initial phase of heavy-ion collisions is that the approaching charged spectators generates a significantly large magnetic field, estimated to be around B~$\sim \; 10^{18}$ Gauss in non-central collisions~\cite{Kharzeev:2007jp,McLerran:2013hla}. Recent model calculations have suggested that such a magnetic field can lead to a substantial difference in $v_{1}(y)$ of positively and negatively charged particles~\cite{Gursoy:2014aka,Gursoy:2016ubw,Gursoy:2018yai}. Experimental efforts have been undertaken at both RHIC and LHC to measure the charge-dependent directed flow of charm hadrons $D^{0}$ and $\overline{D^{0}}$ with the anticipation that charm hadrons, being produced early in the collision process, are more sensitive to the initial large-strength magnetic field than the light quark species~\cite{Das:2016cwd,Chatterjee:2018lsx}. However, a conclusive evidence regarding the charge splitting of $v_{1}$ of charm hadrons has not yet been achieved due to large experimental uncertainties~\cite{Adam:2019wnk,Acharya:2019ijj}. Recently, the STAR collaboration reported a significant charge splitting in $dv_{1}/dy$ ($\Delta dv_{1}/dy$) considering light quark species~\cite{STAR:2023jdd}. In central collisions, a positive $\Delta dv_{1}/dy$ has been observed for protons, kaons, and pions. However, $\Delta dv_{1}/dy$ turns distinctly negative in peripheral collisions. While the positive sign has been linked to the contribution from transported quarks~\cite{Guo:2012qi}, the significant negative sign linked to the influences from initial electromagnetic fields with the Faraday induction dominates over the Coulomb effect. In another publication, STAR also observed that $\Delta dv_{1}/dy$ increases as function of electrical charge difference and strangeness difference in mid-central collisions~\cite{STAR:2023wjl}. Taking into account the combined findings, the $v_{1}$ splittings are interpreted as a competition between the Hall effect and the Faraday induction + Coulomb effect, its flavor and centrality dependence. However, a recent relativistic dissipative fluid dynamic model with baryon diffusion~\cite{Parida:2023ldu,Parida:2023rux} can accommodate the pattern and sign of $\Delta dv_{1}/dy$ for baryon species observed by STAR, but it can not capture the same observed in pions and kaons. In an environment rich in baryons, it is anticipated that various mean-field potentials could induce a splitting effect. When subjected to an attractive potential, particles are inclined to become trapped within the system and move perpendicular to the participant plane. On the other hand, under repulsive potentials, particles are more likely to exit the system and move parallel to the participant plane. Previous studies have explored the influence of such mean-field approaches on the splitting in elliptic flow~\cite{Xu:2013sta}. Nevertheless, further investigation is needed to assess the impact of this mean-field approach on $\Delta dv_{1}/dy$.

Furthermore, the production mechanism of light nuclei, such as deuterons, in heavy-ion collisions is highly debated. Primarily, there are two commonly employed approaches for explaining deuteron production mechanism:  i) the thermal model, in which the production of light nuclei occurs throughout the evolution of the fireball until chemical freeze-out, through interactions involving elementary nucleons or partons~\cite{Mekjian:1978zz,Braun-Munzinger:1994zkz,Chatterjee:2014ysa}; such models are successful in explaining deuteron yields~[\cite{Andronic:2010qu,Cleymans:2011pe}]; ii) the coalescence model, where the formation of light nuclei takes place at a later stage in the collision, near kinetic freeze-out. In this model, two nucleons that are close in proximity and traveling with similar velocities can come together to form a light nucleus~\cite{Butler:1963pp,Gutbrod:1976zzr,Sato:1981ez,Steinheimer:2012tb}. As a result, the momentum distribution of these newly formed light nuclei is closely linked to that of protons. Nevertheless, understanding the formation of light nuclei, such as deuterons, during the intermediate phases of these collisions and their persistence in the subsequent evolution poses a significant challenge. This challenge stems from the fact that their binding energy is relatively modest, typically a few MeV, in contrast to the considerably higher fireball temperature, in the range of hundreds of MeV. Various experiments conducted at AGS and RHIC have investigated the directed flow of various light nuclei species~\cite{EOS:1994kku,EOS:1994jzn,FOPI:2011aa}. Recently, the STAR collaboration reported observations of the $v_{1}$ for light nuclei  and hyper nuclei species across a range of Beam Energy Scan collider and fixed target energies, spanning from 3.0 to 39 GeV~\cite{STAR:2020hya,STAR:2021ozh,STAR:2022fnj}. Several coalescence models have predicted that light nuclei, such as deuterons, behavior should exhibit an atomic mass scaling. However, when it comes to measurements between 7.7 and 39 GeV, no definitive conclusion has been reached due to large uncertainties in the STAR measurements. Interestingly, at 3 GeV where the experimental precision is modest, the STAR experiment observed that all light and hyper-nuclei species adhere to a mass number scaling.

\begin{figure*}[h!]
\centerline{\includegraphics[scale=0.9]{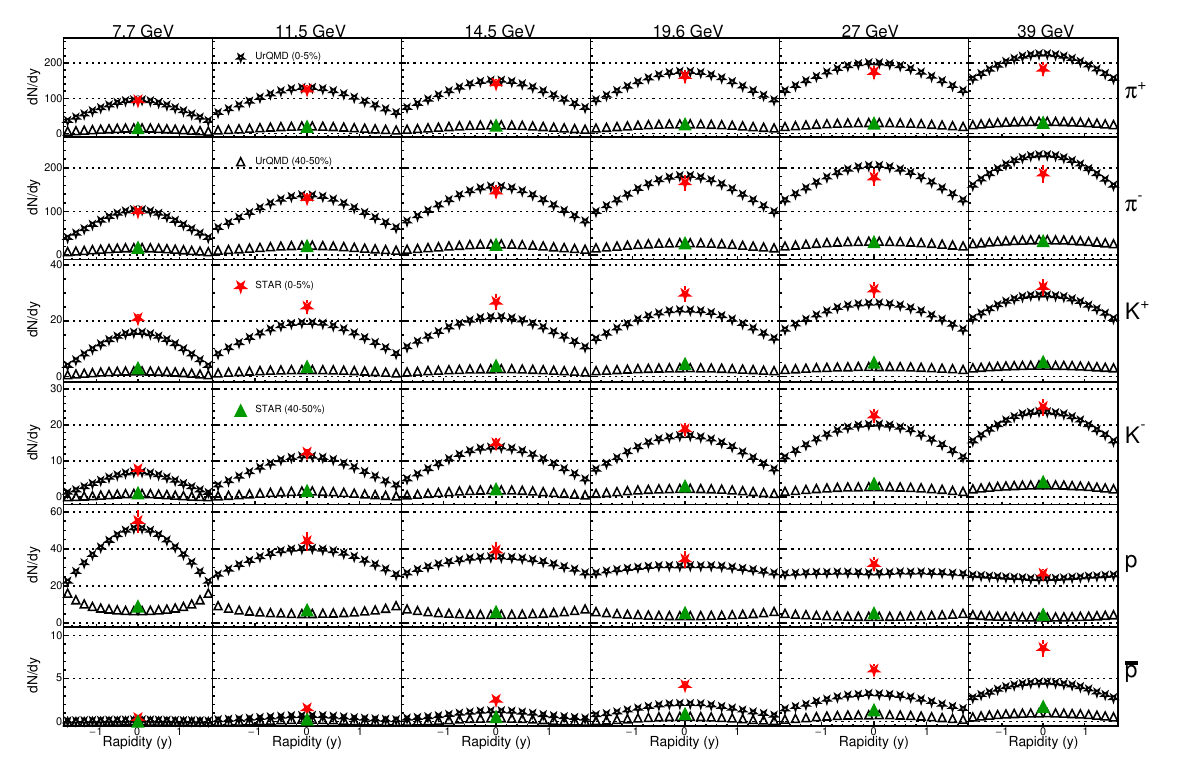}}
\vspace*{8pt}
\caption{Rapidity distribution for $\pi^{\pm}$, $K^{\pm}$, p, $\bar{\mathrm p}$ in 0-5\% and 40-50\% Au+Au collisions at $\sqrt{s_{NN}}$ = 7.7, 11.5, 14.5, 19.6, 27 and 39 GeV using UrQMD with $\tau$ = 20 fm/c. STAR measurements are taken from~\cite{STAR:2017sal, STAR:2019vcp}}
\label{fig0:dndy}
\end{figure*}

In the given context, a comprehensive understanding of the $v_{1}(y)$ of several particle species and its charge-dependence is of utmost importance. It is worth noting that the impact on $v_{1}$ arising from hadronic interactions in the final particle state is frequently disregarded. Several measurements at RHIC-BES have shown good indication of the presence of a substantial hadronic phase. Such as the deviation from quark scaling in $v_{n}$ coefficients~\cite{STAR:2017okv,STAR:2021yiu} and the centrality-dependent ratios of $K^{*0}/K^{-}$ at BES energies favors the dominance of hadronic interactions~\cite{STAR:2022sir}. However, the lifetime of the hadronic phase and its consequence on experimental observables is often neglected. The UrQMD (Ultra Relativistic Quantum Molecular Dynamics) framework~\cite{Bass:1998ca} is based upon a microscopic theory of particle transport. Often this framework is used emulates hadronic interactions as an afterburner. In this paper, we have employed the UrQMD to investigate the influence of late-stage hadronic interactions on the rapidity-odd directed flow, as well as associated charge splitting. The UrQMD model offers the flexibility to tune the $"tim"$ parameter ($\tau$) within the hadronic cascade, thereby controlling the duration of the hadronic phase. By setting $\tau$ to a specific value, an instantaneous freeze-out scenario is simulated. On the other hand, increasing $\tau$ prolongs the interactions among the generated particles, allowing an extended period of hadronic interactions. This treatment provides a means to understand the impact of hadronic interactions across varying time scales of system evolution, and it has recently been utilized to gain insights into the production of resonances at RHIC-BES energies~\cite{Sahoo:2023rko,Sahoo:2023dkv}. 

\begin{figure*}[h!]
\centerline{\includegraphics[scale=0.9]{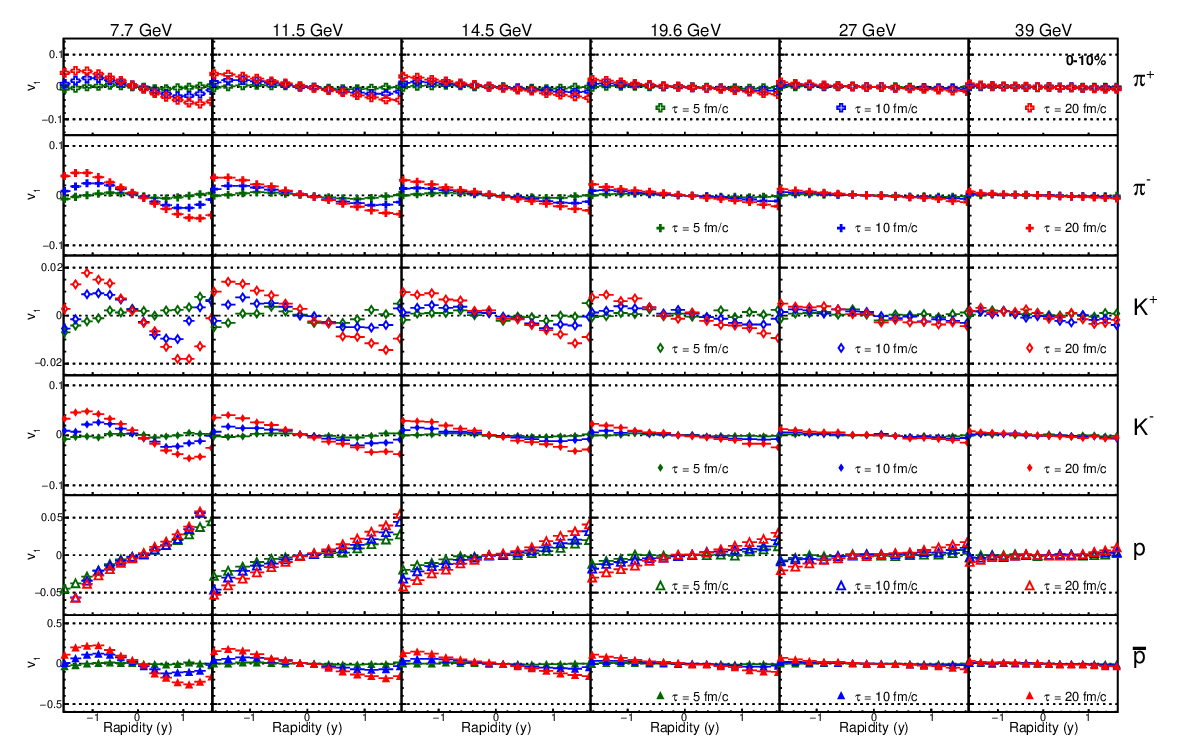}}
\vspace*{8pt}
\caption{Odd-component of directed flow ($v_{1}$) as a function rapidity (y) for $\pi^{\pm}$, $K^{\pm}$, p, $\bar{\mathrm p}$ in 0-10\% Au+Au collisions at $\sqrt{s_{NN}}$ = 7.7, 11.5, 14.5, 19.6, 27 and 39 GeV. Different colors correspond to different lifetimes of the hadronic lifetime parameter $\tau$=5, 10 and 20 fm/c.}
\label{fig1:v1y010}
\end{figure*}

\begin{figure*}[h!]
\centerline{\includegraphics[scale=0.9]{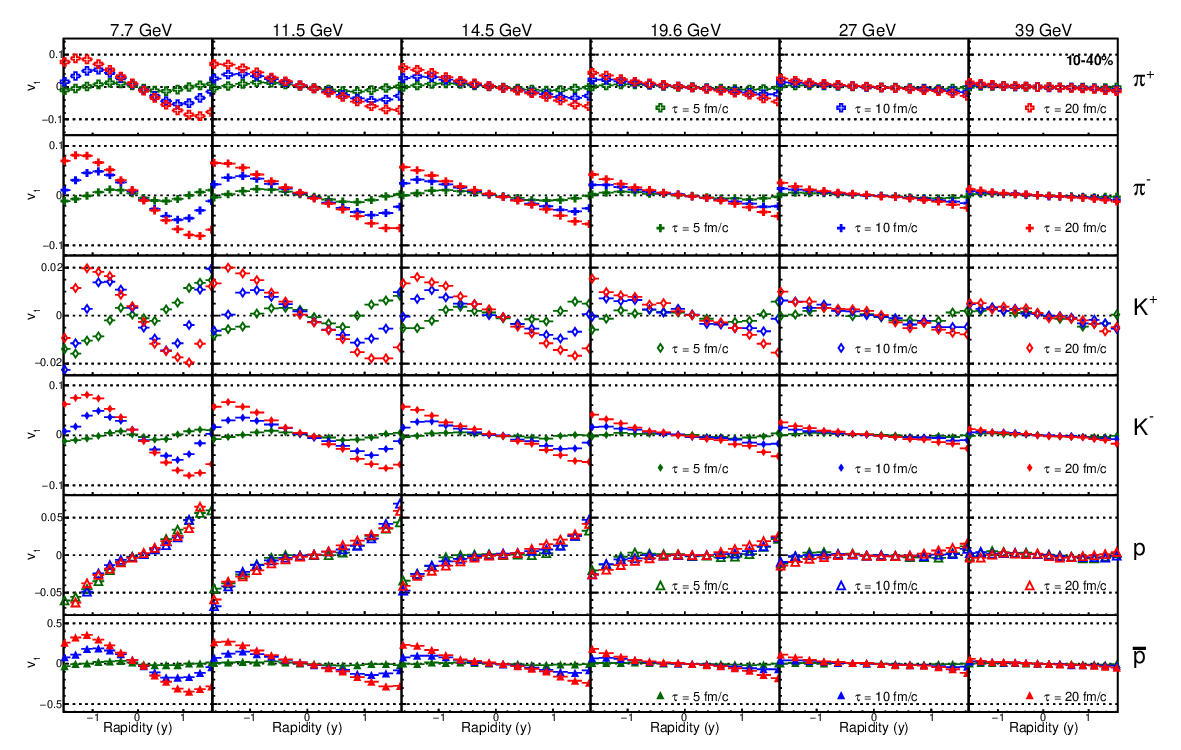}}
\caption{Odd-component of directed flow ($v_{1}$) as a function rapidity (y) for $\pi^{\pm}$, $K^{\pm}$, p, $\bar{\mathrm p}$ in 10-40\% Au+Au collisions at $\sqrt{s_{NN}}$ = 7.7, 11.5, 14.5, 19.6, 27 and 39 GeV. Different colors correspond to different lifetimes of the hadronic lifetime parameter $\tau$=5, 10 and 20 fm/c.}
\label{fig2:v1y1040}
\end{figure*}

\begin{figure*}[h!]
\centerline{\includegraphics[scale=0.9]{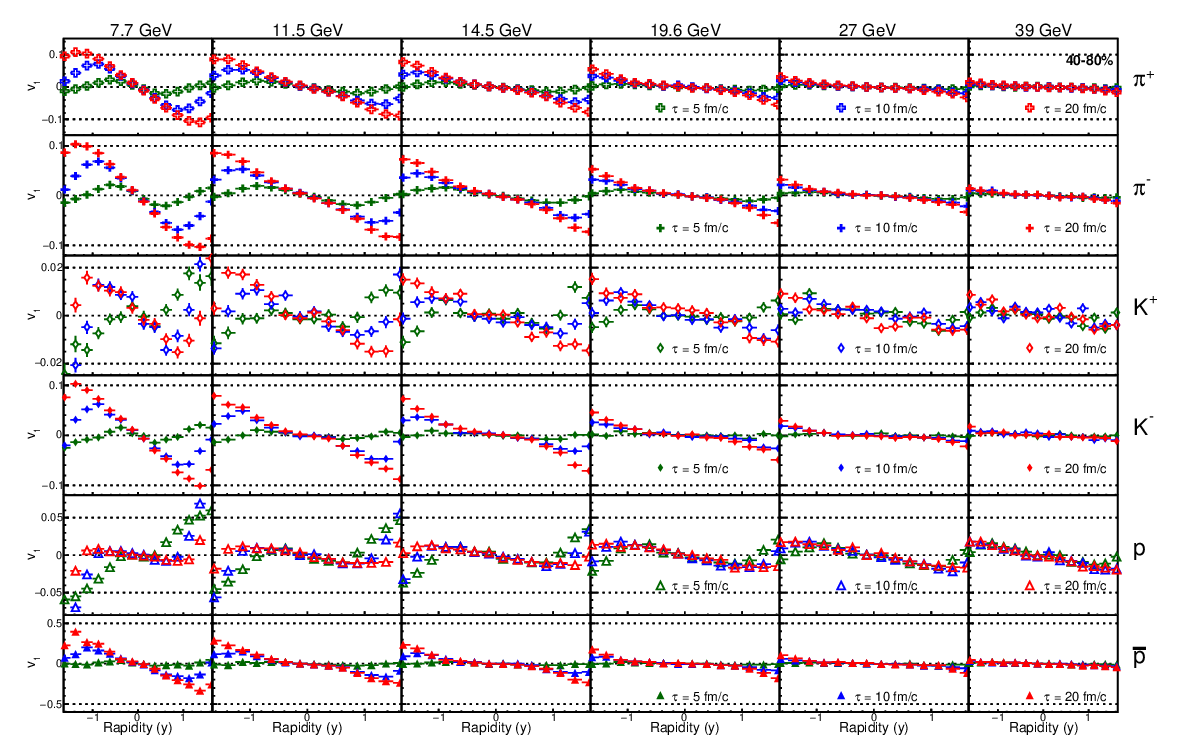}}
\caption{Odd-component of directed flow ($v_{1}$) as a function rapidity (y) for $\pi^{\pm}$, $K^{\pm}$, p, $\bar{\mathrm p}$ in 40-80\% Au+Au collisions at $\sqrt{s_{NN}}$ = 7.7, 11.5, 14.5, 19.6, 27 and 39 GeV. Different colors correspond to different lifetimes of the hadronic lifetime parameter $\tau$=5, 10 and 20 fm/c.}
\label{fig3:v1y4080}
\end{figure*}

\begin{figure*}[h!]
\centerline{\includegraphics[scale=0.7]{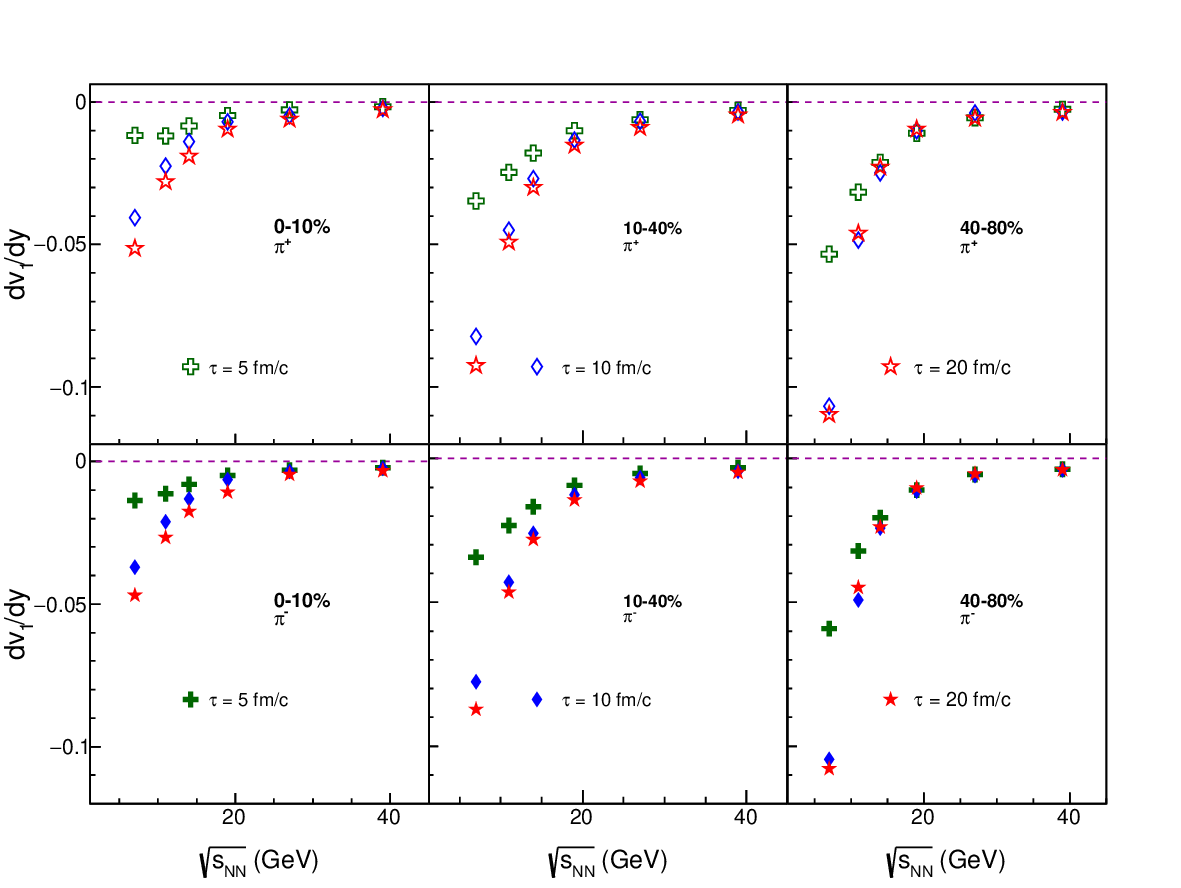}}
\caption{Slope of rapidity dependent $v_{1}$ ($dv_{1}/dy$) as a function of beam energy for $\pi^{+}$ (top panel) and $\pi^{-}$ (bottom panel) with a different lifetime of the hadronic lifetime parameter $\tau$=5, 10 and 20 fm/c.}
\label{fig4:piondv1dy}
\end{figure*}

\begin{figure*}[h!]
\centerline{\includegraphics[scale=0.7]{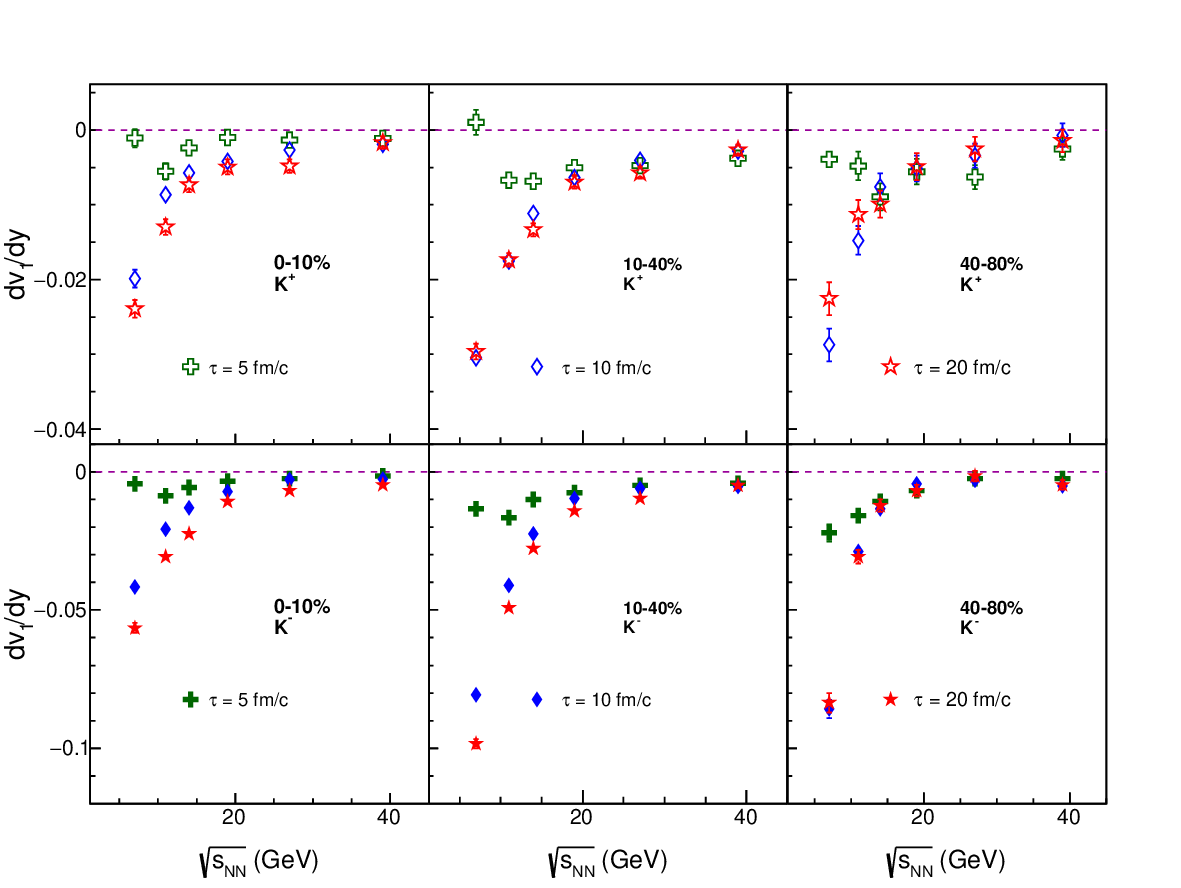}}
\caption{Slope of rapidity dependent $v_{1}$ ($dv_{1}/dy$) as a function of beam energy for $K^{+}$ (top panel) and $K^{-}$ (bottom panel) with a different lifetime of the hadronic lifetime parameter $\tau$=5, 10 and 20 fm/c.}
\label{fig5:kaondv1dy}
\end{figure*}

\begin{figure*}[h!]
\centerline{\includegraphics[scale=0.7]{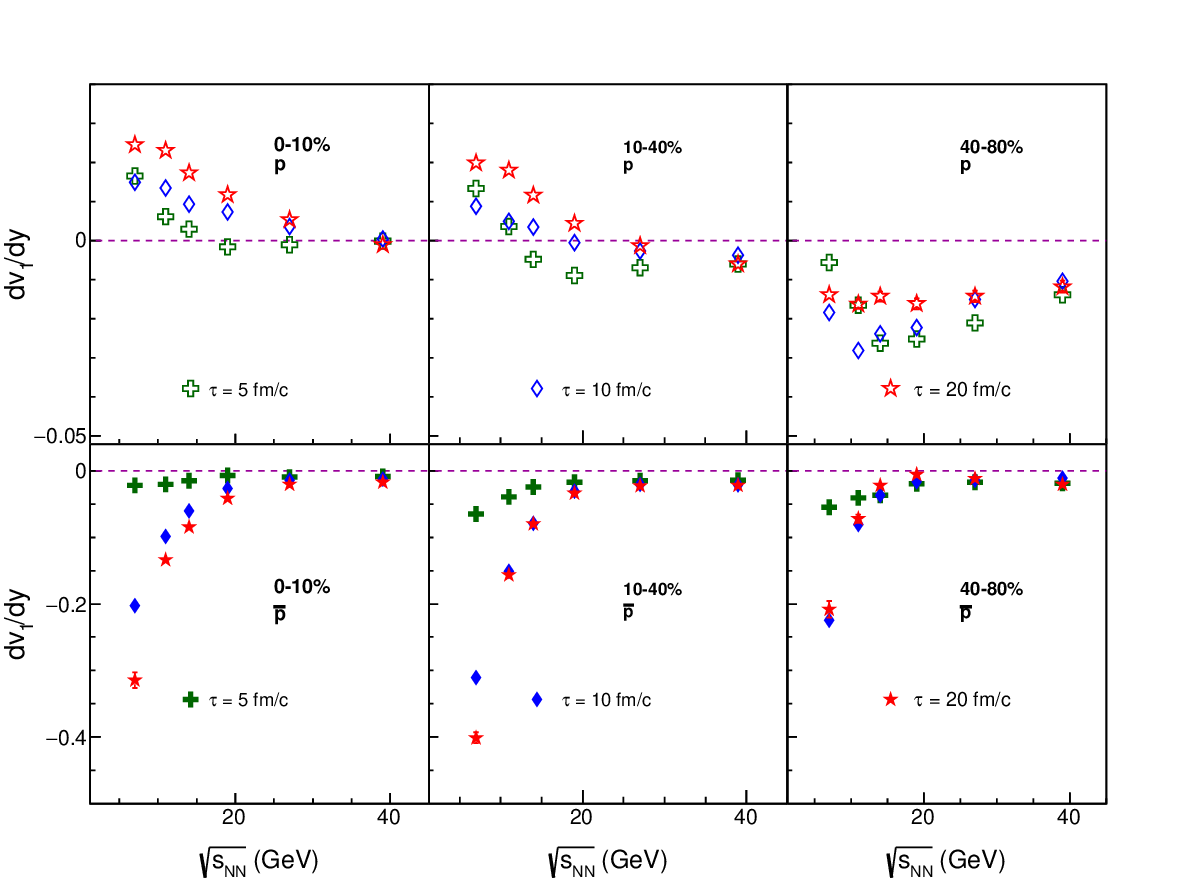}}
\caption{Slope of rapidity dependent $v_{1}$ ($dv_{1}/dy$) as a function of beam energy for $p$ (top panel) and $\bar{p}$ (bottom panel) with a different lifetime of the hadronic lifetime parameter $\tau$=5, 10 and 20 fm/c.}
\label{fig6:protondv1dy}
\end{figure*}

\section{The UrQMD model}
\label{sec-2}
The UrQMD (Ultra Relativistic Quantum Molecular Dynamics) framework is based on a microscopic transport theory. This model enables the simulation of classical trajectories for all types of hadrons, accompanied by stochastic binary scattering, the formation of color strings, and the decay of resonances. The theory encompasses an array of interactions, including baryon-baryon, meson-baryon, and meson-meson interactions. Notably, its collisional component accounts for over 50 species of baryons and 45 species of mesons.\\ 
%Within the model, we varied hadronic cascade $\tau$ from 5 to 50 fm/c. 
In this study, we employed the publicly available UrQMD model. We opted for the default Equation of State (EoS) within UrQMD, known as the CASCADE mode, which involves no potentials (EoS=0). Within this model, resonance masses are distributed according to the Breit-Wigner function, facilitating the simulation of their decay processes. We performed the hadronic simulation or hadron propagation for a duration of $\tau$, specifically choosing values from 5-20 fm/c. Throughout the simulation process, we enabled the decay of unstable particles, allowing us to effectively capture their natural decay mechanisms. We extracted relevant data about the final particles present during the freeze-out stage, from the "test.f14" file, which was generated by the model. Notably, our calculation does not consider phase transitions or electromagnetic fields into account. But it is valuable for establishing a baseline understanding. \
Note that the variation of $\tau$ is used to illustrate how an observable is changing at different time scales in a prolonged hadronic phase. When comparing experimental measurements, we use $\tau$ = 20 fm/c. Furthermore, we have verified that both $dN/dy$ and $v_{1}$ do not change appreciably when changing $\tau$ from 20 to 50 fm/c at RHIC BES energies.

To study the light nuclei, namely deuteron, we have utilized a dynamical coalescence model~\cite{coal-1,coal-2,coal-3}. In this particular model, we utilize phase-space distribution of protons and neutrons at the freezeout to form deuterons. The number of deuteron produced is given by the overlap of phase space distribution function of protons and neutrons ($f_{p,n}(x,k)$) at freeze-out with the Wigner phase space function $\rho_{w}((x_{1} -x_{2}), (k_{1}-k_{2})/2)$ inside the deuteron, as shown in Eq.~\ref{eq-2}.\\

\begin{equation}
\label{eq-2}
    \begin{split}
        N_{d} = g_{d} \int{d^{3}\boldsymbol{x_{1}}}\int{d^{3} \boldsymbol{k_{1}}}\int{d^{3}\boldsymbol{x_{2}}}\int{d^{3}\boldsymbol{k_{2}}}f_{n}(\boldsymbol{x_{1},k_{1}})f_{p}
    (\boldsymbol{{x_{2}},k_{2}}) \times \\ \rho_{w}((\boldsymbol{x_{1}} -\boldsymbol{x_{2}}), (\boldsymbol{k_{1}}-\boldsymbol{k_{2}})/2)
    \end{split}
  \end{equation}
 Here, $g_{d}$ implies the statistical factor for deuteron, with its value set at 3/8. This choice is made by considering the combined spin and isospin of the proton and neutron, as detailed in references~\cite{Mattiello:1996gq,coal-3}. It's crucial to note that conflicting values for $g_{d}$ exist in the literature. For example, references~\cite{Bond:1977fd,Sato:1981ez} use a $g_{d}$ value of 3/4, which is also employed to describe deuteron yield in~\cite{Sun:2017xrx,Sun:2022xjr}. However, it's evident that the selection of $g_{d}$ has an impact only on the deuteron count and not on the magnitude of directed flow. $\boldsymbol{x_{i}}$ and $\boldsymbol{p_{i}}$ are the position and momentum coordinates of the nucleons after freezeout obtained from UrQMD model. Intitally $\boldsymbol{x_{i}}$ and $\boldsymbol{p_{i}}$ are in the center of mass frame of two colliding nucleus. A Lorentz transformation is performed to get these coordinates in the rest frame of deuteron. To perform coalescence at equal times in the rest frame of the deuteron, nucleons that freeze out earlier are permitted to freely move with a constant velocity. This velocity is determined by their respective energies and momenta, and they continue to move until the last nucleon in the deuteron cluster also undergoes freezeout. The Wigner distribution function inside deuteron is given by Eq.~\ref{Eq-3}

\begin{equation}
\label{Eq-3}
    \rho_{w} = 8\exp{(-r^{2}/\sigma_{d}^{2} - \sigma_{d}^{2}k^{2})}.
\end{equation}
Here, $r$ = $|{\boldsymbol{r_{1}} - \boldsymbol{r_{2}}}|$ and $k$ = $|({\boldsymbol{k_{1}} -\boldsymbol{k_{2}}})/2|$
are the relative position and momentum coordinates of the two nucleons. The parameter $\sigma_{d}$ is related to the RMS charge radius of the deuteron, $\sigma_{d} = \sqrt{8/3}r_{d}$. In this study, we have taken $\sigma_{d}$ = 3.2 fm corresponds to the RMS charge radius of 1.96 fm.  We have additionally confirmed that varying the $\sigma_{d}$ parameter within a reasonable range (50\%) has no significant impact on directed flow.
 
\section{UrQMD Results}
First, we calculate the rapidity distributions for 
$\pi^{\pm}$, $K^{\pm}$, p, and $\bar{\mathrm p}$ in Au+Au collisions at center-of-mass energies ($\sqrt{s_{NN}}$) of 7.7, 11.5, 14.5, 19.6, 27, and 39 GeV using UrQMD with $\tau$ = 20 fm/c. It is presented in Fig~\ref{fig0:dndy} for 0-5\% and 40-50\% collision centralities. The UrQMD calculation can reasonably explain the mid-rapidity yield of identified hadrons measured by STAR~\cite{STAR:2017sal, STAR:2019vcp}. Figures \ref{fig1:v1y010}, \ref{fig2:v1y1040}, and \ref{fig3:v1y4080} illustrate the rapidity dependence of $v_{1}$ for various particle species including $\pi^{\pm}$, $K^{\pm}$, p, and $\bar{\mathrm p}$ in Au+Au collisions at center-of-mass energies ($\sqrt{s_{NN}}$) of 7.7, 11.5, 14.5, 19.6, 27, and 39 GeV and spanning centrality ranges of 0-10\%, 10-40\%, and 40-80\%. Each subplot employs distinct markers in different colors to represent outcomes obtained with different $\tau$ values, namely 5, 10, and 20 fm/c.

The magnitude of $v_{1}$ increases as beam energy decreases, in line with the expectation of more stopping near mid-rapidity. Additionally, an increase in $v_{1}$ is apparent with a larger value of $\tau$. Typically the strength of $v_{1}$ is quantified by its slope with respect to rapidity (called $dv_{1}/dy$). To quantify this, a fitting procedure employs the function $p_{0}y + p_{1}y^{3}$, where $p_{0}$ equates to $dv_{1}/dy$. This functional form effectively captures the rapidity-odd behavior of $v_{1}(y)$ and its bending in the forward region.

Focusing on Figures \ref{fig4:piondv1dy}, \ref{fig5:kaondv1dy}, and \ref{fig6:protondv1dy}, the upper and lower panels respectively depict the energy-dependent variation of $dv_{1}/dy$ for $\pi^{\pm}$, $K^{\pm}$, and p ($\bar{\mathrm p}$). $\pi^{+}$ and $\pi^{-}$ exhibit a consistently negative $dv_{1}/dy$ across the 7.7--39 GeV range, with the slope becoming more pronounced as the hadronic phase time $\tau$ increases. Similar trends, akin to pions, are observed for $K^{+}$ and $K^{-}$, with the exception of a tendency towards positive $dv_{1}/dy$ for $K^{+}$ at $\sqrt{s_{NN}}$ = 7.7 GeV. It is possibly suggesting contribution of associated production ($pp \rightarrow p \Lambda (1115) K^{+}$), where the proton is strongly correlated with the $K^{+}$~\cite{Balewski:1998pd,Zhou:2017jfk}.
Notably, the energy-dependent pattern of protons and antiprotons is starkly different from that of pions and kaons. In 0-10\% collisions, protons predominantly display positive $v_{1}$ values, whereas antiprotons consistently exhibit negative values for all $\tau$ cases. The occurrence of sign reversal in $v_{1}$ is observed to vary with the duration of the hadronic phase, both in 0-10\% and 10-40\% centrality ranges. Meanwhile, for peripheral collisions, both protons and antiprotons exhibit negative $dv_{1}/dy$. Note that the sign of proton $v_{1}$ is often linked with the signature of softening in the equation of state associated with a first-order phase transition. The observed shift in the position of sign change in $v_{1}$ slope with varying hadronic cascade time underscores the necessity of accounting for hadronic evolution and its duration when interpreting the sign reversal in proton $v_{1}$.

\begin{figure*}[h!]
\centerline{\includegraphics[scale=0.9]{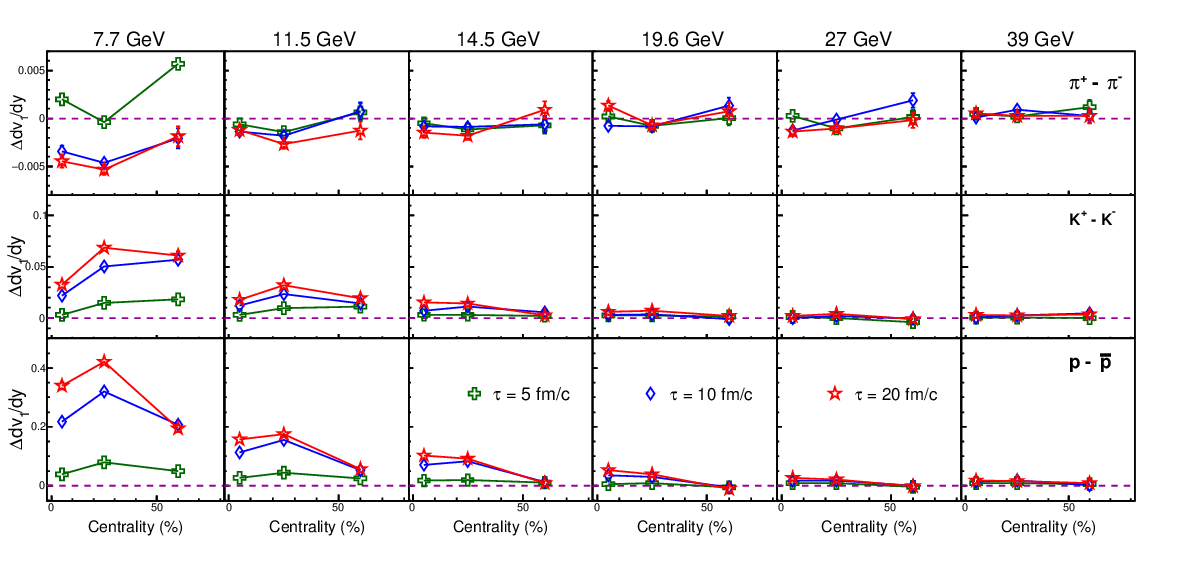}}
\caption{Difference in directed flow rapidity slope ($\Delta dv_{1}/dy$) of positively and negatively charged $\pi$, K and p as a function of centrality for Au+Au collisions at $\sqrt{s_{NN}}$ = 7.7, 11.5, 14.5, 19.6, 27 and 39 GeV with a different lifetime of the hadronic lifetime parameter $\tau$=5, 10 and 20 fm/c.}
\label{fig7:deltav1dy}
\end{figure*}

Multiple theoretical calculations have pointed that the quarks originating from the initial-state nuclei (such as $u$ and $d$ quarks in proton) acquire different $v_{1}$ than quarks produced via pair production (such as $\bar{u}$ and $\bar{d}$ quarks in anti-proton)~\cite{Dunlop:2011cf,Goudarzi:2020eoh}. The constituent quarks originating from such different sources could give rise to differences in the $v_{1}$ for positively and negatively charged hadrons~\cite{Wang:2018pqx}. Figure \ref{fig7:deltav1dy} depicts the difference of $dv_{1}/dy$ (referred to as $\Delta dv_{1}/dy$) between positively and negatively charged hadrons across different centrality bins for collision energies ranging from $\sqrt{s_{\mathrm NN}}$ = 7.7 to 39 GeV. It is noteworthy that the magnitude of $\Delta dv_{1}/dy$ increases as the beam energy decreases, encompassing all the particle species under investigation. For pions, this difference tends to hover around zero when $\sqrt{s_{NN}}$ surpasses 19.6 GeV, while at lower energies, it tends to shift towards the negative side. At 7.7 GeV, $\Delta dv_{1}/dy$ exhibits a marked dependence on centrality across all particle species. This dependence on centrality is particularly strong for central and mid-central collisions, where the magnitude of $\Delta dv_{1}/dy$ escalates with an increase in $\tau$ from 5 to 20 fm/c. Conversely, peripheral collisions show a relatively smaller change in $\Delta dv_{1}/dy$ across different $\tau$ values. Our observation indicates that a prolonged hadronic phase, expected in central collisions, enhances the magnitude of $\Delta dv_{1}/dy$. While in peripheral collisions, $\Delta dv_{1}/dy$ is not altered significantly by hadronic interactions. In UrQMD, the above effect is more prominent at 7.7 GeV.

\begin{figure*}[h!]
\centerline{\includegraphics[scale=0.9]{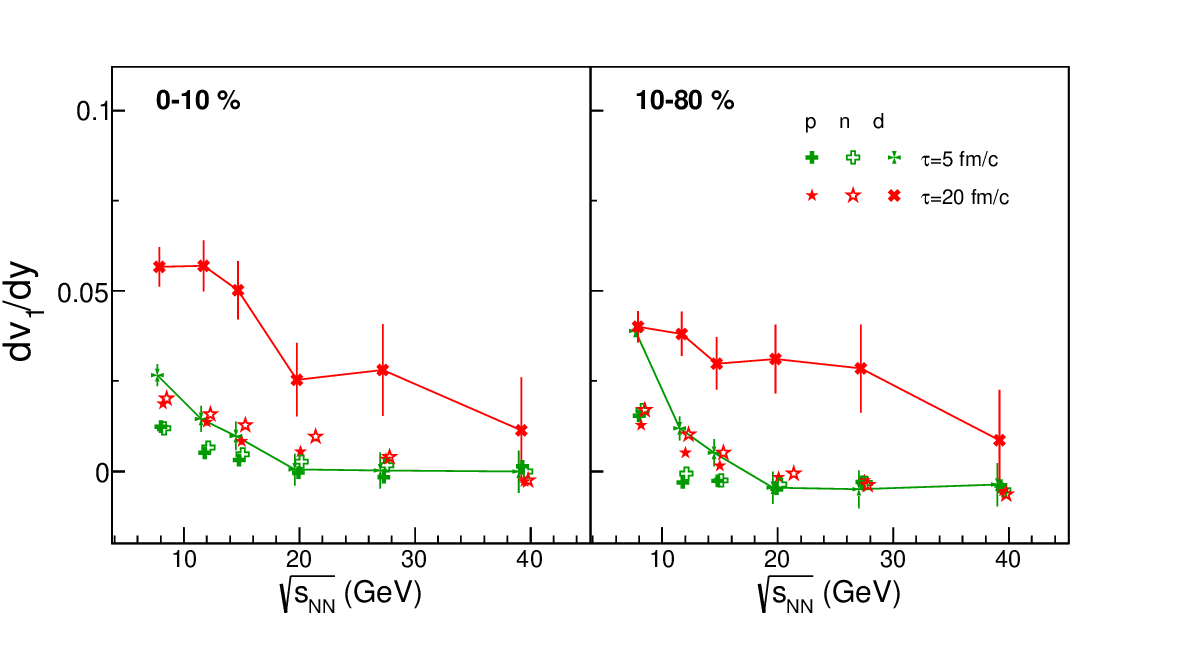}}
\caption{Beam energy dependence of $dv_{1}/dy$ for proton, neutron and deuteron with different hadronic lifetime parameter $\tau$=5 and 20 fm/c.}
\label{fig8:dv1dy_deuteron}
\end{figure*}

To understand the impact of hadronic interactions on light nuclei, which is expected to be formed at a later stage of the collisions, we have studied the directed flow of deuterons at $\sqrt{s_{NN}}$ = 7.7 - 39 GeV. The deuterons are formed via phase space coalscence of protons and neutrons as described in section~\ref{sec-2}. The figure~\ref{fig8:dv1dy_deuteron} depicts the beam energy dependence of the mid-rapidity $v_{1}$ slope ($dv_{1}/dy$) of deuterons for different hadronic cascade lifetime parameters $\tau$=5 and 20 fm/c. These results are compared to protons and neutrons for the same $\tau$ parameter. It is observed that with increasing $\tau$, the $dv_{1}/dy$ of d are also increasing. The relative increase of d $dv_{1}/dy$ with $\tau$ is much larger compared to the increase in $dv_{1}/dy$ of p and n at a same energy. Note that in our modeling, late-stage interactions have a direct impact on the p and n, whereas the d's are generated from the nucleons at the end of the freeze-out process. As a result, the combined influence of late-stage interactions on p and n is magnified through the coalescence mechanism, subsequently reflected in the d.

%Therefore the combined effect of late-stage interaction of p and n is getting amplified due to coalscence mechanism and reflected on the deuterons.

\section{Comparison to experimental measurements}
\begin{figure*}[h!]
\centerline{\includegraphics[scale=0.7]{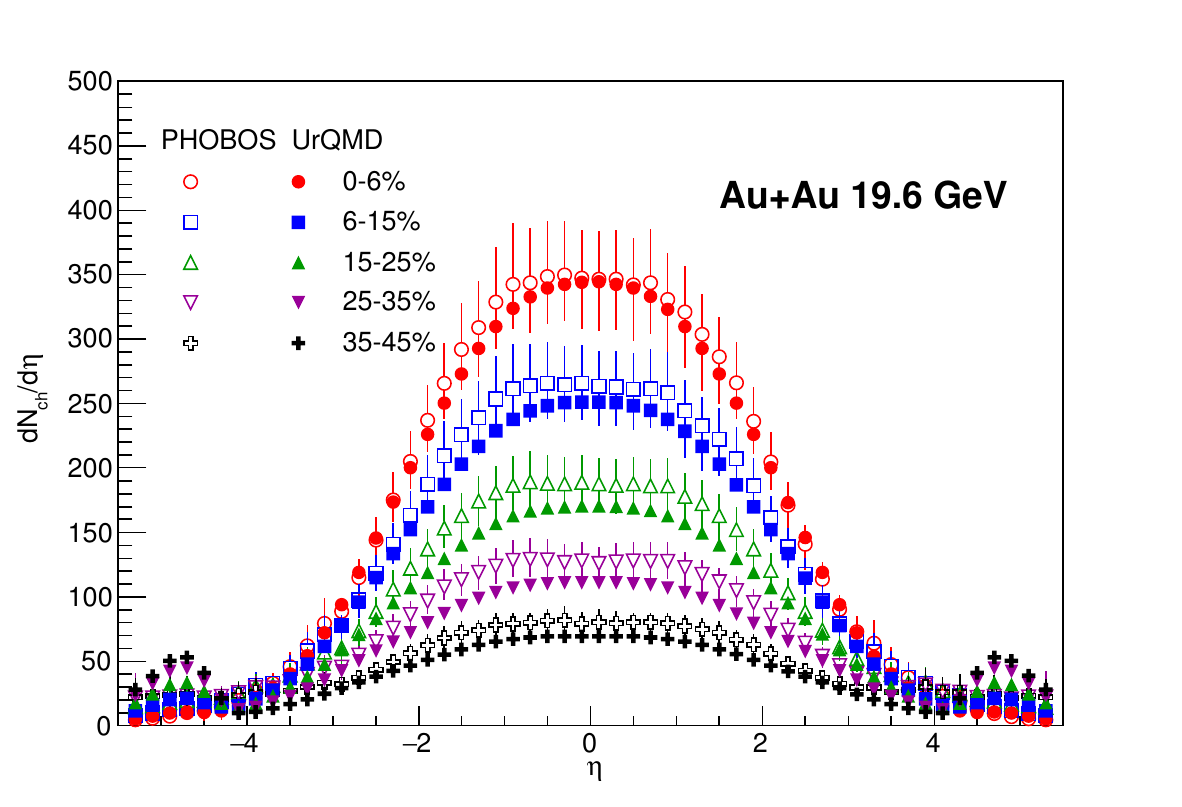}}
\caption{Comparison of  $dN_{ch}/d\eta$ for inclusive charged hadrons in Au+Au collisions at $\sqrt{s_{NN}}$=19.6 GeV by PHOBOS~\cite{Back:2002wb} with UrQMD calculations with hadronic lifetime parameter $\tau$ = 20 fm/c.}
\label{fig10:dndeta_data_model}
\end{figure*}

\begin{figure*}[h!]
\centerline{\includegraphics[scale=0.7]{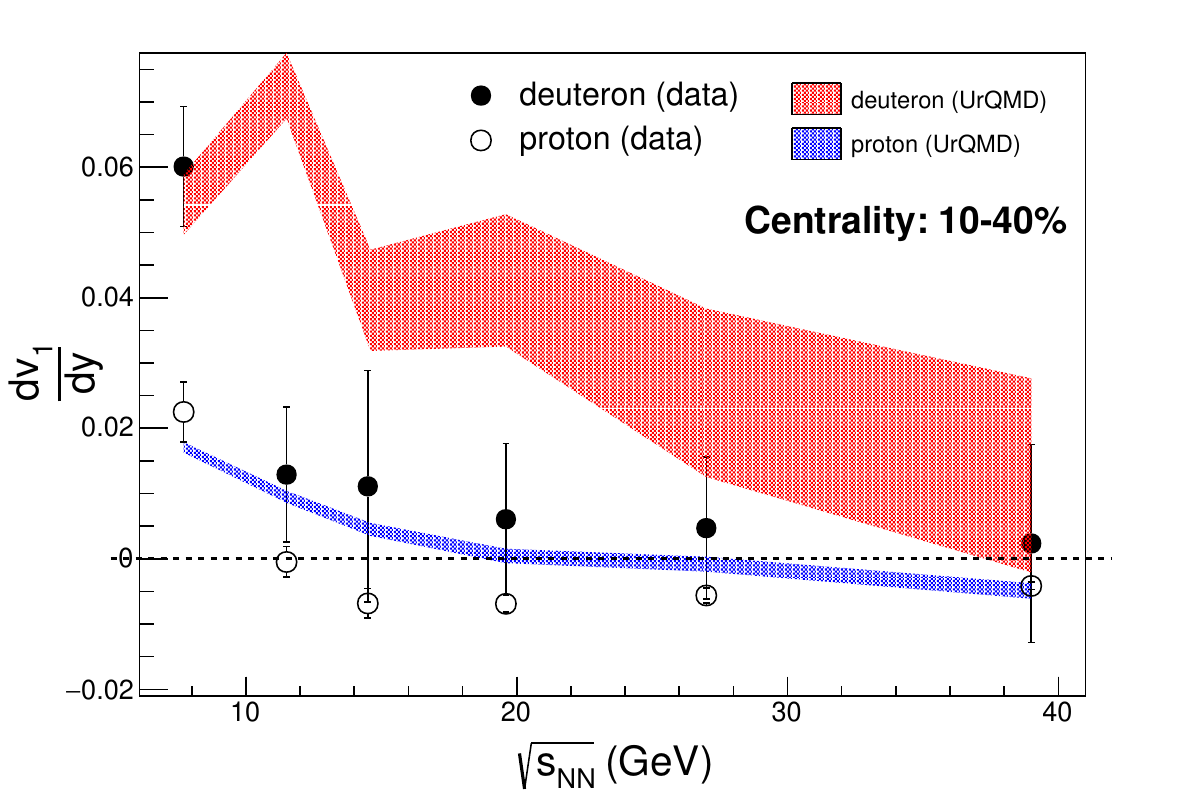}}
\caption{Comparison of  directed flow of protons and deuterons from Beam Energy Scan Au+Au collisions by STAR~\cite{STAR:2014clz, STAR:2020hya} with UrQMD calculations with hadronic lifetime parameter $\tau$ = 20 fm/c.}
\label{fig11:v1_data_model}
\end{figure*}

In this section, we compare our UrQMD calculations with some existing experimental measurements from RHIC. Illustrated in Figure~\ref{fig10:dndeta_data_model}, our UrQMD results, employing a $\tau$ value of 20 fm/c, is consistent with the charged particle pseudorapidity distribution measured by PHOBOS in 19.6 GeV Au+Au collisions~\cite{Back:2002wb}. Notably, the UrQMD roughly captures the observed data across a broad rapidity spectrum and various collision centralities. 
Figure~\ref{fig11:v1_data_model} presents a comparison of the energy dependence of $dv_{1}/dy$ for protons and deuterons, as measured by STAR BES~\cite{STAR:2014clz, STAR:2020hya}, with UrQMD using $\tau$ = 20 fm/c. Notably, the proton data from STAR exhibits a sign change at approximately 10 GeV, whereas the UrQMD simulation indicates a sign change occurring at a significantly higher energy, around 19 GeV. In the case of deuterons simulated with UrQMD employing coalescence, a positive sign for $dv_{1}/dy$ is observed. However, it tends to overestimate the corresponding STAR data. Additional studies required to achieve a quantitative description of data. 
\begin{figure*}[h!]
\centerline{\includegraphics[scale=0.7]{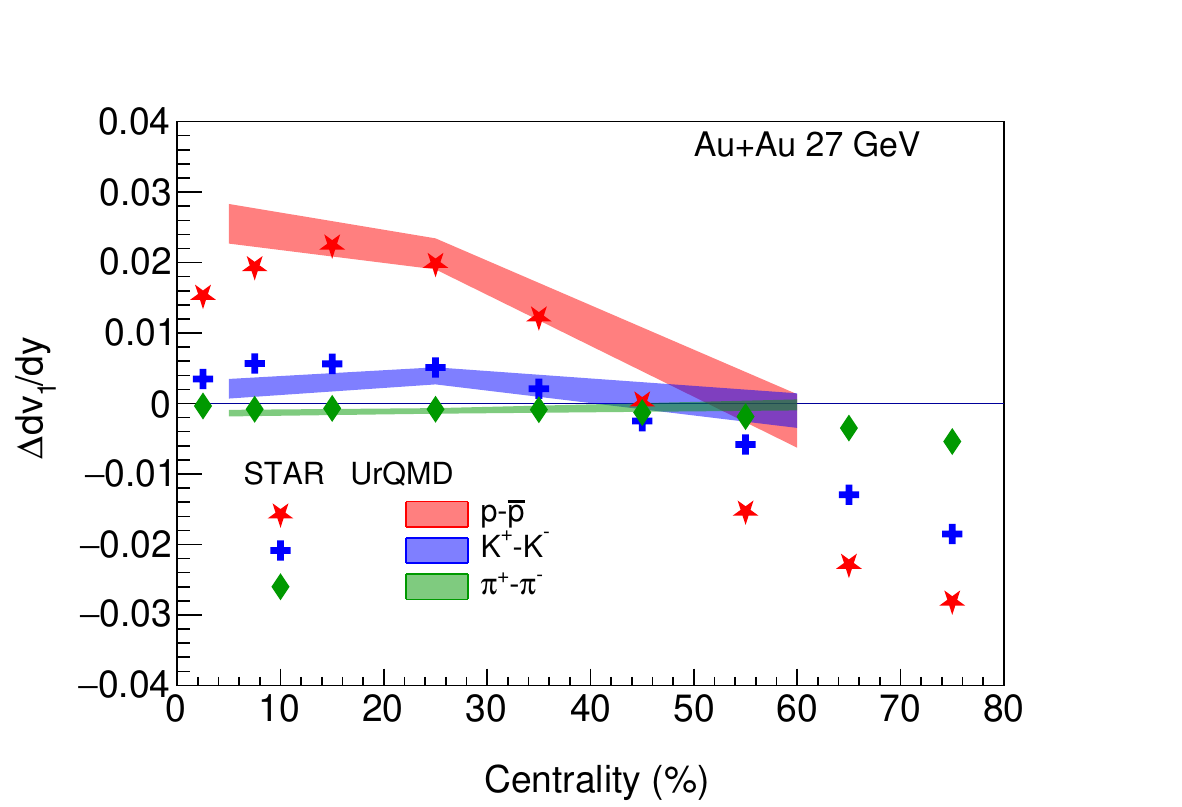}}
\caption{Comparison of $\Delta dv_{1}/dy$ for $\pi^{\pm}$, $K^{\pm}$ and p ($\bar{p}$) for 27 GeV Au+Au collisions by STAR~\cite{STAR:2023jdd} with UrQMD calculations with $\tau$ = 20 fm/c.}
\label{fig12:Dellta_v1_data_model}
\end{figure*}
The Figure~\ref{fig12:Dellta_v1_data_model} illustrates a comparison of the charge splitting for $\pi^{\pm}$, $K^{\pm}$, and p ($\bar{p}$) at 27 GeV Au+Au, as measured by STAR~\cite{STAR:2023jdd}, alongside UrQMD calculations with $\tau$ = 20 fm/c. The rich structure evident in the $\Delta dv_{1}/dy$ values of the measured particle species can not be explained by UrQMD. In particular, at 27 GeV, UrQMD predicts negligible $v_{1}$ charge splitting for pions and kaons, in contrast to the observed data. However, it is worth noting that the positive $\Delta dv_{1}/dy$ observed for protons in UrQMD can be attributed to baryon stopping along with contributions from late-stage hadronic interactions.

\section{Conclusion}
We presented a comprehensive study of rapidity density distribution ($dN/dy$), rapidity dependent of odd-directed flow ($v_1$) and its slope near mid-rapidity ($dv_{1}/dy$) in Au+Au collisions at $\sqrt{s_{\mathrm NN}}$ = 7.7 to 39 GeV from UrQMD model. To illustrate the role of prolonged hadronic phase, we varied hadronic cascade lifetime ($\tau$). The manner in which $dv_{1}/dy$ for p ($\bar{\mathrm p}$) changes with energy exhibits a distinctive pattern when compared to $\pi^{\pm}$ and $K^{\pm}$. It is worth noting that there is an observed shift in the position of sign reversal in proton $dv_{1}/dy$ at varying beam energies, along with changing $\tau$, in both central and mid-central collisions. Furthermore, the centrality-dependent difference in $dv_{1}/dy$ between positively and negatively charged hadrons ($\Delta dv_{1}/dy$) displays a noticeable variation for different types of particles. Interestingly, the deuterons, which are formed through the nucleon coalescence mechanism, exhibit a notable increase in their $dv_{1}/dy$ as the parameter $\tau$ increases. This study emphasizes the significance of accounting for the duration of the hadronic phase while interpreting the sign change in $v_{1}$ slopes and its charge dependence as well as the light nuclei formation in heavy ion collisions. Furthermore, we compared some available experimental data on $dN/dy$, $dv_{1}/dy$, and $\Delta dv_{1}/dy$ with UrQMD calculations at $\tau = 20$ fm/c. While UrQMD can capture certain qualitative features of the data, it falls short in providing a quantitative explanation. More investigations are needed to understand the roles of the equation of state, mean field, baryon stopping, and electromagnetic field in directed flow.

\section*{Acknowledgements}

The authors acknowledge discussions with Sandeep Chatterjee and Tribhuban Parida. AKS and SS acknowledges support from the Strategic Priority Research Program of the Chinese Academy of Sciences (Grant No. XDB34000000). 

%% The Appendices part is started with the command \appendix;
%% appendix sections are then done as normal sections
%\appendix

%\bibliographystyle{apsrev4-1}
\bibliographystyle{ws-mpla} 
\bibliography{references}
\end{document}